\documentclass[a4paper]{article}

\usepackage[ninepoint]{itgspeech2018}    
\usepackage{times}            
\usepackage[english]{babel}   
\usepackage[utf8]{inputenc}
\usepackage[T1]{fontenc}      
\usepackage[sort&compress,numbers]{natbib}	
\usepackage{amsmath,amssymb}
\usepackage{graphicx}
\usepackage[colorlinks=false,pdfborder={0 0 0}]{hyperref}
\usepackage{units}

\usepackage[nolist]{acronym}
\usepackage{siunitx}
\usepackage{bm}
\usepackage[subtle, mathspacing=tight, wordspacing=tight, bibbreaks=normal, floats=normal, tracking=normal, paragraphs=tight]{savetrees}

\newcommand\mythanks[1]{%
  \begingroup
  \renewcommand\thefootnote{}\footnote{#1}%
  \addtocounter{footnote}{-1}%
  \endgroup
}

\title{RTF-Based Binaural MVDR Beamformer Exploiting an External Microphone in a Diffuse Noise Field}

\author{Nico G\"o\ss ling, Simon Doclo}

\address{\small University of Oldenburg, Department of Medical Physics and Acoustics and Cluster of Excellence Hearing4All, Oldenburg, Germany\\
  Email: \texttt{\{nico.goessling, simon.doclo\}@uni-oldenburg.de}\\
  Web: \texttt{www.sigproc.uni-oldenburg.de}}
  
\newcommand{\Rn}{\mathbf{R}_{\rm n}}
\newcommand{\Rx}{\mathbf{R}_{\rm x}}
\newcommand{\Ry}{\mathbf{R}_{\rm y}}
\newcommand{\vv}[1]{\mathbf{#1}}
\newcommand{\vy}{\vv{y}}
\newcommand{\vx}{\vv{x}}
\newcommand{\vn}{\vv{n}}
\newcommand{\va}{\vv{a}}
\newcommand{\val}{\vv{a}_{\rm L}}
\newcommand{\var}{\vv{a}_{\rm R}}
\newcommand{\vel}{\vv{e}_{\rm L}}
\newcommand{\ver}{\vv{e}_{\rm R}}
\newcommand{\vwl}{\vv{w}_{\rm L}}
\newcommand{\vwr}{\vv{w}_{\rm R}}
\newcommand{\phixl}[0]{\phi_{\rm x,L}}
\newcommand{\phixr}[0]{\phi_{\rm x,R}}

\begin{document}
\ninepointfont
\sloppy
%
\begin{acronym}
	\acro{atf}[ATF]{acoustic transfer function}
	\acro{rtf}[RTF]{relative transfer function}
	\acro{mvdr}[BMVDR]{binaural minimum variance distortionless response beamformer}
	\acro{psd}[PSD]{power spectral density}
	\acro{gevd}[GEVD]{generalized eigenvalue decomposition}
	\acro{sc}[SC]{spatial coherence}
	\acro{snr}[SNR]{signal-to-noise ratio}
	\acro{cw}[CW]{covariance whitening}
\end{acronym}
\maketitle
\begin{abstract}
Besides suppressing all undesired sound sources, an important objective of a binaural noise reduction algorithm for hearing devices is the preservation of the binaural cues, aiming at preserving the spatial perception of the acoustic scene.
A well-known binaural noise reduction algorithm is the binaural minimum variance distortionless response beamformer, which can be steered using the relative transfer function (RTF) vector of the desired source, relating the acoustic transfer functions between the desired source and all microphones to a reference microphone.
In this paper, we propose a computationally efficient method to estimate the RTF vector in a diffuse noise field, requiring an additional microphone that is spatially separated from the head-mounted microphones.
Assuming that the spatial coherence between the noise components in the head-mounted microphone signals and the additional microphone signal is zero, we show that an unbiased estimate of the RTF vector can be obtained.
Based on real-world recordings, experimental results for several reverberation times show that the proposed RTF estimator outperforms the widely used RTF estimator based on covariance whitening and a simple biased RTF estimator in terms of noise reduction and binaural cue preservation performance.
\end{abstract}
%
%
\section{Introduction}\label{sec:intro}
%
\mythanks{This work was supported by the Collaborative Research Centre 1330
Hearing Acoustics. the Cluster of Excellence 1077 Hearing4all, funded
by the German Research Foundation (DFG), and by the joint Lower Saxony-Israeli Project ATHENA.}
Noise reduction algorithms for head-mounted assistive listening devices (e.g., hearing aids, cochlear implants, hearables) are crucial to improve speech intelligibility and speech quality in noisy environments.
Binaural noise reduction algorithms are able to use the spatial information captured by all microphones on both sides of the head \cite{Doclo2015,Doclo2018}.
Besides suppressing undesired sound sources, binaural noise reduction algorithms also aim at preserving the listener's spatial perception of the acoustic scene to assure spatial awareness, to reduce confusions due to a possible mismatch between acoustical and visual information, and to enable the listener to exploit the binaural hearing advantage \cite{Bronkhorst1988}.\\
As shown in \cite{Cornelis2010,Doclo2015,Doclo2018}, the \ac{mvdr} beamformer is able to preserve the binaural cues, i.e., the interaural level difference (ILD) and the interaural time difference (ITD), of the desired source.
The \ac{mvdr} beamformer can either be implemented using the \acp{atf} between the desired source and all microphones or using the \acp{rtf}, relating the \acp{atf} to a reference microphone \cite{Gannot2001}.
Since estimating the RTFs (unlike the ATFs) is feasible in practice, RTF estimation has become an important task in the field of multichannel speech enhancement \cite{Cohen2004,Warsitz2007,Markovich2009,Krueger2011,Serizel2014,Markovich2015,Giri2016,Varzandeh2017}.\\
Aiming at improving the performance of (binaural) noise reduction algorithms, recently the use of an external microphone in combination with the head-mounted microphones has been explored \cite{Bertrand2009,Cvijanovic2013,Szurley2016,Farmani2017,Goessling2017,Goessling2017b,Yee2017J,Ali2018}.
It has, e.g., been shown that using an external microphone is able to improve performance in terms of noise reduction \cite{Bertrand2009,Szurley2016,Goessling2017,Goessling2017b,Yee2017J,Ali2018}, source localisation \cite{Farmani2017} and binaural cue preservation \cite{Szurley2016,Goessling2017}.\\
In this paper, we propose a computationally efficient method to estimate the RTF vector in a diffuse noise field using the external microphone.
This method requires the external microphone to be located far enough from the head-mounted microphones, such that the spatial coherence between the noise components in the head-mounted microphone signals and the external microphone signal is low.
Assuming this spatial coherence to be zero, we show how an unbiased RTF estimator can be derived.
Using real-world recordings, we compare the proposed RTF estimator to a simple biased RTF estimator and to the widely used RTF estimator based on \ac{cw} \cite{Markovich2009,Warsitz2007,Krueger2011,Serizel2014,Markovich2015} for several reverberation times and \acp{snr}.
The results show that the proposed RTF estimator yields a larger SNR improvement and reduced binaural cue errors compared to the existing RTF estimators.
When comparing the proposed RTF estimator to an oracle RTF estimator (using the clean speech signal as external microphone signal), only a small performance difference can be observed.
%
\section{Configuration and Notation}\label{sec:config}
%
\begin{figure}[t]
	\setlength{\unitlength}{\linewidth}
	\begin{picture}(1,0.6)
		\put(0.25,0){\includegraphics[width=0.4\linewidth]{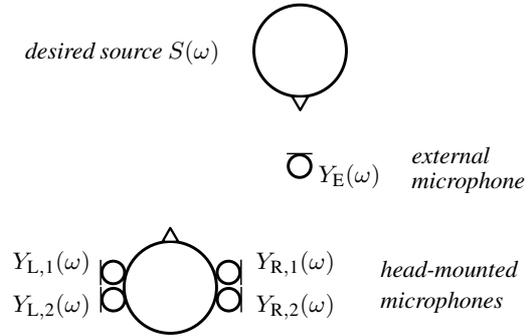}}
		\put(0.6,0.25){$Y_{\rm{E}}(\omega)$}
		\put(0.13,0.45){\it{desired source} $S(\omega)$}
		\put(0.11,0.11){$Y_{{\rm L},1}(\omega)$}
		\put(0.11,0.04){$Y_{{\rm L},2}(\omega)$}
		\put(0.5,0.11){$Y_{{\rm R},1}(\omega)$}
		\put(0.5,0.04){$Y_{{\rm R},2}(\omega)$}
		\put(0.7,0.097){\it head-mounted}
		\put(0.7,0.047){\it microphones}
		\put(0.75,0.28){\it external}
		\put(0.75,0.24){\it microphone}
	\end{picture}
	\caption{Top-view of the considered acoustic scenario and microphone configuration ($M=2$).}
	\label{fig:setup}
\end{figure}
We consider an acoustic scenario with one desired source $S(\omega)$ and diffuse background noise (e.g., cylindrically or spherically isotropic noise) in a reverberant enclosure.
Moreover, we consider a binaural configuration, consisting of a left and a right device (each containing $M$ microphones), and an external microphone that is spatially separated from the head-mounted microphones, cf. Figure 1.
The $m$-th microphone signal of the left hearing device $Y_{{\rm{L}},m}(\omega)$ can be written in the frequency-domain as
\begin{equation}\label{eq:Y}
  Y_{{\rm{L}},m}(\omega) = X_{{\rm{L}},m}(\omega) + N_{{\rm{L}},m}(\omega) \, , \quad m \in \{1,\dots,M\},
\end{equation}
where $X_{{\rm L},m}(\omega)$ denotes the desired speech component, $N_{{\rm L},m}(\omega)$ denotes the noise component and $\omega$ denotes the angular frequency. For conciseness we will omit $\omega$ in the remainder of the paper, wherever possible. The $m$-th microphone signal of the right hearing device $Y_{{\rm{R}},m}$ and the external microphone $Y_{{\rm{E}}}$ are similarly defined by substituting $\rm R$ and $\rm E$ for $\rm L$, respectively.
The microphone signals of the hearing devices can be stacked in a vector, i.e.,
\begin{equation}\label{eq:y}
  \vy = \left[ Y_{{\rm L},1}, \; \dots,\; Y_{{\rm L},M}, \; Y_{{\rm R},1}, \; \dots,\; Y_{{\rm R},M} \right]^T \in \mathbb{C}^{2M}\, ,
\end{equation}
with $(\cdot)^T$ denoting the transpose of a vector.
Using \eqref{eq:Y}, the vector $\vy$ can be written as
\begin{equation}
  \vy = \vx + \vn \, ,
\end{equation}
where the speech vector $\vx$ and the noise vector $\vn$ are defined similarly as in \eqref{eq:y}.
Without loss of generality, we choose the first microphone on each hearing device as reference microphone, i.e.,
\begin{equation}
  Y_{\rm L} = \vel^T\vy \, , \quad Y_{\rm R} = \ver^T\vy \, ,
\end{equation}
where $\vel$ and $\ver$ are selection vectors consisting of zeros and one element equal to 1, i.e., $\vel(1) = 1$ and $\ver(M+1) = 1$.
In the case of a single desired source, the speech vector $\vx$ is equal to
\begin{equation}\label{eq:x}
  \vx = \va S \, ,
\end{equation}
where the vector $\va \in \mathbb{C}^{2M}$ contains the \acp{atf} between the desired source $S$ and all microphones, including reverberation, microphone characteristics and head-shadowing.
The \ac{rtf} vectors $\val$ and $\var$ of the desired source are defined by relating the \ac{atf} vector $\va$ to both reference microphones, i.e.,
\begin{equation}\label{eq:rtfvec}
  \val = \frac{\va}{\vel^T\va} \, , \quad \var = \frac{\va}{\ver^T\va} \, .
\end{equation}
The speech covariance matrix $\Rx \in \mathbb{C}^{2M\times 2M}$ and the noise covariance matrix $\Rn \in \mathbb{C}^{2M \times 2M}$ are defined as
\begin{align}
\label{eq:Rx}
	\Rx &= \mathcal{E}\{\vx\vx^H\} = \phixl\val\val^H = \phixr\var\var^H \, ,\\
	\label{eq:Rn}
	\Rn &= \mathcal{E}\{\vn\vn^H\} \, ,
\end{align} 
where $\mathcal{E}\{\cdot\}$ denotes the expectation operator, $(\cdot)^H$ denotes the conjugate transpose, and $\phixl = \mathcal{E}\{|X_{\rm L}|^2\}$ and $\phixr = \mathcal{E}\{|X_{\rm R}|^2\}$ denote the \ac{psd} of the desired source in the reference microphones.
Assuming statistical independence between the desired speech and noise components, the microphone signal covariance matrix is equal to
\begin{equation}
  \Ry = \mathcal{E}\{\vy\vy^H\} = \Rx + \Rn \, .
\end{equation}
The output signals at the left and the right hearing device are obtained by filtering and summing all microphone signals using the complex-valued filter vectors $\vwl$ and $\vwr$, respectively, i.e.,
\begin{equation}
  Z_{\rm L} = \vwl^H\vy \, , \quad Z_{\rm R} = \vwr^H\vy \, .
\end{equation}

%
\section{Binaural MVDR Beamformer}\label{sec:mvdr}
%
In this section, we briefly review the well-known BMVDR beamformer \cite{Doclo2010,Doclo2018,Klasen2007}.
The \ac{mvdr} beamformer minimizes the output noise \ac{psd} while preserving the desired speech component in the reference microphones, hence preserving the binaural cues of the desired source.
The constrained optimization problem for the left filter vector is given by
\begin{equation}
  \min_{\vwl} \; \mathcal{E}\{|\vwl^H\vn|^2\} \quad \text{subject to} \quad \vwl^H\val=1 \, .
\end{equation}
The constrained optimization problem for the right filter vector is defined similarly by substituting $\rm R$ for $\rm L$.
The solutions of these optimization problems are equal to \cite{Doclo2015,Doclo2018,Gannot2001}
\begin{align}\label{eq:mvdr}
  \vwl = \frac{\Rn^{-1}\val}{\val^H\Rn^{-1}\val} \, , \quad \vwr = \frac{\Rn^{-1}\var}{\var^H\Rn^{-1}\var} \, .
\end{align}
Hence, to calculate the BMVDR beamformer an estimate of the noise covariance matrix $\Rn$ and the \ac{rtf} vectors $\val$ and $\var$ of the desired source is required.
Usually, the noise covariance matrix $\Rn$ is either estimated by recursively updating the matrix during speech pauses or approximated by using an appropriate model, e.g., assuming a spherically isotropic noise field.
Similarly, the RTF vectors $\val$ and $\var$ are either estimated from the microphone signals or approximated by using -- simulated or measured -- anechoic RTFs corresponding to the assumed position of the desired source (e.g., in front of the user).
In the following sections we will consider data-dependent RTF estimation approaches to steer the \ac{mvdr} beamformer in \eqref{eq:mvdr}.
%
\section{RTF Estimation Approaches}\label{sec:rtf}
%
In this section, we discuss different approaches to estimate the \ac{rtf} vectors $\val$ and $\var$ of the desired source.
First, we consider a biased estimator, which only requires an estimate of the microphone signal covariance matrix $\Ry$.
Second, we consider the CW estimator \cite{Markovich2009,Serizel2014}, which requires estimates of the microphone signal covariance matrix $\Ry$ and the noise covariance matrix $\Rn$.
Third, we present an RTF estimator that exploits the external microphone signal $Y_{\rm E}$, assuming the spatial coherence between the noise components in the head-mounted microphone signals and the external microphone signal is zero.
\subsection{Biased Estimator (B)}
Using \eqref{eq:rtfvec} and \eqref{eq:Rx}, it can be easily shown that the RTF vectors are equal to
\begin{equation}\label{eq:rtftrue}
  \val = \frac{\Rx\vel}{\vel^T\Rx\vel} \, , \quad \var = \frac{\Rx\ver}{\ver^T\Rx\ver} \, ,
\end{equation}
i.e., a column of the speech covariance matrix $\Rx$ normalized with the element corresponding to the respective reference microphone. When no reliable estimate of the speech covariance matrix $\Rx$ is available, a simple but biased RTF estimate can be obtained by using the (noisy) microphone signal covariance matrix $\Ry$ \cite{Braun2015}
\begin{equation}\label{eq:B}
  \val^{\rm B} = \frac{\Ry\vel}{\vel^T\Ry\vel} \, , \quad \var^{\rm B} = \frac{\Ry\ver}{\ver^T\Ry\ver} \, .
\end{equation}
The biased estimator in \eqref{eq:B} obviously does not lead to the same solution as \eqref{eq:rtftrue}, especially for low input \acp{snr}.
\subsection{Covariance Whitening (CW)}
A frequently used (unbiased) RTF estimator is based on covariance whitening \cite{Markovich2009,Warsitz2007,Krueger2011,Serizel2014,Markovich2015}.
Using a square-root decomposition (e.g., Cholesky decomposition), the noise covariance matrix $\Rn$ can be written as
\begin{equation}
  \Rn = \Rn^{H/2}\Rn^{1/2} \, .
\end{equation}
The pre-whitened microphone signal covariance matrix is then equal to
\begin{equation}
  \Ry^{\rm w} = \Rn^{-H/2}\Ry\Rn^{-1/2} \, ,
\end{equation}
which can be decomposed using the eigenvalue decomposition (EVD) as
\begin{equation}
  \Ry^{\rm w} = \mathbf{V}\mathbf{\Lambda}\mathbf{V}^H \, ,
\end{equation}
where the matrix $\mathbf{V} \in \mathbb{C}^{2M \times 2M}$ contains the eigenvectors and the diagonal matrix $\mathbf{\Lambda} \in \mathbb{R}^{2M \times 2M}$ contains the corresponding eigenvalues. Using the principal eigenvector $\mathbf{v}_{\rm max}$, i.e., the eigenvector corresponding to the largest eigenvalue, the RTF vectors can be estimated as \cite{Markovich2015}
\begin{equation}\label{eq:CW}
  \val^{\rm CW} = \frac{\Rn^{1/2}\mathbf{v}_{\rm max}}{\vel^T\Rn^{1/2}\mathbf{v}_{\rm max}} \, , \quad \var^{\rm CW} = \frac{\Rn^{1/2}\mathbf{v}_{\rm max}}{\ver^T\Rn^{1/2}\mathbf{v}_{\rm max}} \, .
\end{equation}
Due to the EVD, this estimator has a larger computational complexity than the biased estimator. Additionally, an estimate of both the microphone signal covariance matrix $\Ry$ and the noise covariance matrix $\Rn$ is required, although this estimate is required anyway for the BMVDR beamformer, cf. \eqref{eq:mvdr}.
\subsection{Spatial Coherence (SC)}
\begin{figure}[t]
  \includegraphics[width=\linewidth]{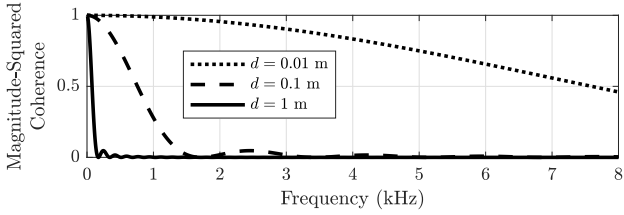}
  \caption{Analytical inter-microphone magnitude-squared coherence in a spherically isotropic noise field.}
  \label{fig:sinc}
\end{figure}
Considering a spherically isotropic noise field as an example for a diffuse noise field, the magnitude-squared coherence (MSC) between the noise components in two different microphones (neglecting head-shadowing) is equal to \cite{Cron1962}
\begin{equation}
  {\rm MSC} = \left|{\rm sinc}\left(\frac{\omega d}{c}\right)\right|^2 \, ,
\end{equation}
where $d$ denotes the distance between the two microphones and $c$ denotes the speed of sound.
Figure \ref{fig:sinc} depicts the MSC for $d \in \left\{0.01, \; 0.1, \; 1 \right\}\si{m}$ and $c = \SI{343}{\meter\per\second}$.
It can be seen that for large distances between the microphones the MSC tends to be very small, especially for high frequencies.\\
For now, let us assume that the external microphone is sufficiently far away from the head-mounted microphones, such that
\begin{equation}\label{eq:noiseCorr}
  \mathcal{E}\{\vn N_{\rm E}^*\} = \mathbf{0} \, ,
\end{equation}
i.e., the noise components in the head-mounted microphone signals are spatially uncorrelated with the noise component in the external microphone signal. Using \eqref{eq:noiseCorr} yields
\begin{equation}\label{eq:corrY}
  \mathcal{E}\{\vy Y_{\rm E}^*\} = \mathcal{E}\{\vx X_{\rm E}^*\} + \mathcal{E}\{\vn N_{\rm E}^*\} = \mathcal{E}\{\vx X_{\rm E}^*\} \, .
\end{equation}
Using \eqref{eq:corrY} and $\vx = X_{\rm L}\val = X_{\rm R}\var$, the spatial-coherence-based RTF estimator (SC) is equal to
\begin{equation}\label{eq:SC}
\boxed{
  \val^{\rm SC} = \frac{\mathcal{E}\{\vy Y_{\rm E}^* \}}{\mathcal{E}\{Y_{\rm L}Y_{\rm E}^*\}}, \quad \var^{\rm SC} = \frac{\mathcal{E}\{\vy Y_{\rm E}^* \}}{\mathcal{E}\{Y_{\rm R}Y_{\rm E}^*\}}
  }
\end{equation}
Of course, in practice the assumption made in \eqref{eq:noiseCorr} does not perfectly hold.
Hence, in the experimental evaluation in Section 5 we also consider an oracle version of the estimator in \eqref{eq:SC}, which uses the clean speech signal $S$ as the external microphone signal, such that \eqref{eq:noiseCorr} perfectly holds, i.e.,
\begin{equation}\label{eq:SCopt}
   \val^{\rm SC_{opt}} = \frac{\mathcal{E}\{\vy S^* \}}{\mathcal{E}\{Y_{\rm L}S^*\}} \, , \quad \var^{\rm SC_{opt}} = \frac{\mathcal{E}\{\vy S^* \}}{\mathcal{E}\{Y_{\rm R}S^*\}} \, .
\end{equation}
Compared to the CW estimator, the SC estimator does not need an estimate of the noise covariance matrix $\Rn$ and has a lower computational complexity, but obviously requires an external microphone to be available.
%
\section{Experimental Results}\label{sec:exp}
%
In this section, an experimental evaluation is presented of the BMVDR beamformer in \eqref{eq:mvdr} using the RTF estimators discussed in Section 4.
In Section 5.1 the recording setup is described, while detailed information about the implementation is provided in Section 5.2 and the results are presented in Section 5.3.
\subsection{Recording setup}
\begin{figure}[t]
  \includegraphics[width=\linewidth]{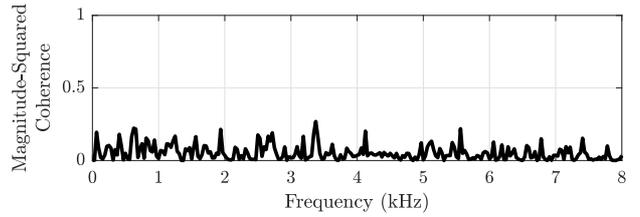}
  \caption{Measured long-term magnitude-squared coherence between the recorded noise in the left reference microphone and the external microphone.}
  \label{fig:noise}
\end{figure}

All signals were recorded in a laboratory located at the University of Oldenburg where the reverberation time can be easily changed by closing and opening absorber panels mounted to the walls and the ceiling.
The room dimensions are about $(7 \times 6 \times 2.7)\; \si{m}$, where the reverberation time was set approximately to the three different values $T_{60} \in \{250, \; 500, \; 750\} \si{ms}$.
The reverberation times were measured using the broad band energy decay curve of measured impulse responses.
At the center of the room a KEMAR head-and-torso simulator (HATS) was placed.
Two behind-the-ear hearing aid dummies with two microphones each, i.e., $M=2$, were placed on the ears of the HATS.\\
The desired source was a male English speaker played back by a loudspeaker placed at about \SI{2}{m} from the center of the head at the same height and at an angle of \SI{35}{\degree}, i.e., on to the right side of the HATS (cf. Figure \ref{fig:setup}).
The external microphone was placed at about \SI{0.5}{\meter} from the desired source, leading to a distance of about \SI{1.5}{\meter} to the HATS, which refers to, e.g., a table microphone or a smartphone that is connected to the binaural hearing device.
To generate the background noise, we used four loudspeakers facing the corners of the laboratory, playing back different multi-talker recordings.
Figure \ref{fig:noise} shows the long-term magnitude-squared coherence between the recorded noise in the reference microphone of the left hearing aid and the external microphone.
It can be observed that the assumption in \eqref{eq:noiseCorr} obviously does not perfectly hold, but the coherence is fairly small.
The desired source and the background noise were recorded separately in order to be able to mix them together at different input ${\rm SNRs} \in \{-5,0,5\}\;\si{dB}$.
The SNR in the external microphone signal was about \SI{9.6}{dB} higher than in the head-mounted microphone signals.
Please note, that streaming and directly using the external microphone signal would not include any binaural cues.
The complete signal had a length of \SI{20}{s} with \SI{0.5}{\second} of noise-only at the beginning.
\begin{figure*}
  \includegraphics[width=\textwidth]{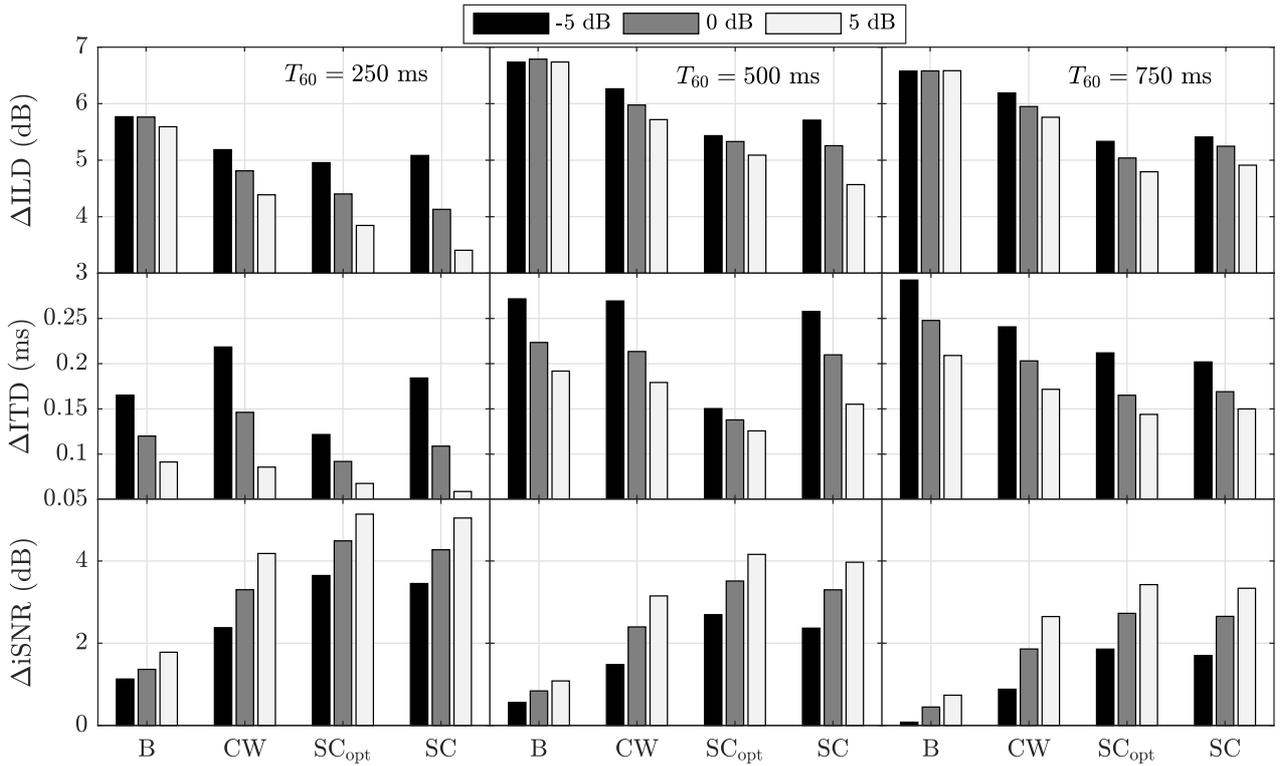}
  \caption{Binaural cue errors and intelligibility-weighted SNR improvement for the RTF estimators for different reverberation times (250 ms, 500 ms, 750 ms) and different input SNRs (-5 dB, 0 dB, 5 dB).}
  \label{fig:results}
\end{figure*}
\subsection{Implementation and Performance Measures}
All signals were processed at a sampling rate of \SI{16}{kHz}. We used the short-time Fourier transform (STFT) with frame length $T=256$, corresponding to \SI{16}{ms}, overlapping by $R = 128$ samples, e.g., for the left reference microphone signal
\begin{align}
  \label{eq:stft}
  Y_{\rm L}(k,l) &= \sum_{t=0}^{T-1} y_{\rm L}(l \cdot R + t)w(t)e^{-j2\pi k t / T} \; ,\\
  &= X_{\rm L}(k,l) + N_{\rm L}(k,l) \, ,
\end{align}
with $k$ the frequency bin index, $l$ the time frame index, $y_{\rm L}(t)$ the left reference microphone signal in the time-domain, $w(t)$ a square-root Hann window of length $T$ and $j = \sqrt{-1}$.\\
To distinguish between speech-plus-noise and noise-only frames we used an oracle broad band voice activity detection (VAD), based on the energy of the speech component in the right reference microphone signal.
Using this VAD, the microphone signal covariance matrix $\hat{\mathbf{R}}_{\rm y}(k,l)$ and the noise covariance matrix $\hat{\mathbf{R}}_{\rm n}(k,l)$ were recursively estimated as
\begin{align}
	\hat{\mathbf{R}}_{\rm y}(k,l) &= \alpha_{\rm y} \hat{\mathbf{R}}_{\rm y}(k,l-1) + (1-\alpha_{\rm y}) \vy(k,l)\vy^H(k,l) \, ,\\
	\hat{\mathbf{R}}_{\rm n}(k,l) &= \alpha_{\rm n} \hat{\mathbf{R}}_{\rm n}(k,l-1) + (1-\alpha_{\rm n}) \vy(k,l)\vy^H(k,l) \, ,
\end{align}
during detected speech-plus-noise frames and noise-only frames, respectively.
The forgetting factors were chosen as $\alpha_{\rm y} = 0.8521$ and $\alpha_{\rm n} = 0.9841$, corresponding to time constants of \SI{50}{ms} and \SI{500}{ms}, respectively.
As initialization the corresponding long-term estimates of the covariance matrices were used.\\
The (time-varying) estimates of the covariance matrices were then used in the biased RTF estimator ($\rm B$) in \eqref{eq:B}, the covariance-whitening-based RTF estimator ($\rm CW$) in \eqref{eq:CW}, the oracle spatial-coherence-based RTF estimator ($\rm SC_{opt}$) in \eqref{eq:SCopt} and the spatial-coherence-based ($\rm SC$) RTF estimator in \eqref{eq:SC}.
We then computed the (time-varying) BMVDR beamformer in \eqref{eq:mvdr} using the estimated RTF vectors and the estimated noise covariance matrix $\hat{\mathbf{R}}_{\rm n}(k,l)$.
The resulting BMVDR beamformer was then applied to the head-mounted microphone signals, i.e.,
\begin{equation}
  Z_{\rm L}(k,l) = \vwl^H(k,l)\vy(k,l) \, , \quad Z_{\rm R}(k,l) = \vwr^H(k,l)\vy(k,l) \, .
\end{equation}
The performance was evaluated in terms of noise reduction and binaural cue preservation.
As a measure for noise reduction performance we used the intelligibility-weighted SNR improvement ($\Delta{\rm iSNR}$) \cite{Greenberg1993} between the right reference microphone signal and the output of the right hearing aid.
As a measure for binaural cue preservation performance we used the reliable binaural cue errors of the direct sound of the desired speech component, i.e., $\rm \Delta ILD$ and $\rm \Delta ITD$, based on an auditory model \cite{Dietz2012} and averaged over frequency.
\subsection{Results}
Figure \ref{fig:results} depicts the results for all four considered RTF estimators for different reverberation times and input SNRs.
As expected, B generally shows worst performance in terms of binaural cue preservation and noise reduction performance.\\
Considering the ILD error, it can be observed for all estimators the ILD errors generally increase for increasing $T_{60}$ and decreasing input SNR.
In addition it can be observed that the SC estimator consistently outperforms the CW estimator, especially for large $T_{60}$.
Moreover, almost no difference can be observed between the SC estimator and the oracle ${\rm SC_{opt}}$  estimator, for all $T_{60}$ and input SNRs.\\
Considering the ITD errors, it can be observed that for all estimators the ITD errors generally increase for increasing $T_{60}$ and decreasing input SNRs.
Contrary to the ILD error, the SC estimator typically leads to larger ITD errors than the oracle ${\rm SC_{opt}}$ estimator, especially for $T_{60} = \SI{250}{ms}$ and \SI{500}{ms}.
Informal listening tests showed that when using SC (and ${\rm SC_{opt}}$) the desired source is perceived as a point source and sounded slightly less reverberated than the input of the reference microphones. For B and CW the binaural cue error sometimes showed large variations over frequency, which may lead to strange sounding artefacts, such that some frequencies are perceived as coming from another direction and the desired source sounds slightly diffuse.\\
Considering the iSNR improvement, it can be observed that for all estimators the SNR improvement generally decreases for increasing $T_{60}$ and decreasing input SNR.
In addition, it can be observed that the SC estimator consistently outperforms the CW estimator for all $T_{60}$ and input SNRs.
Moreover, almost no difference can be observed between the SC estimator and the oracle ${\rm SC_{opt}}$ estimator.
From these results it can be concluded that the SC estimator outperforms the CW estimator.
Moreover, for the considered scenario, i.e., the external microphone about \SI{0.5}{m} from the desired source and about \SI{1.5}{m} from the head-mounted microphones, the overall performance of the (practically implementable) SC estimator is very similar to the oracle ${\rm SC_{opt}}$ estimator, showing that the spatial coherence assumption in \eqref{eq:noiseCorr} is valid for the considered scenario.
It can be expected that placing the external microphone closer to the desired source would slightly improve the performance of the SC estimator, especially in terms of binaural cue preservation.

\section{Conclusions}\label{sec:conclu}
%
In this paper we have shown how an external microphone signal can be exploited to estimate the RTF vectors of a desired source in a diffuse noise field.
We assumed the spatial coherence between the noise components in the head-mounted microphone signals and the noise component in the external microphone signal to be zero to derive an unbiased RTF estimator.
An experimental evaluation using real-world signals for several reverberation times and input SNRs showed that a better noise reduction performance and binaural cue preservation can be obtained when using the proposed RTF estimator compared to an RTF estimator based on covariance whitening and a simple biased RTF estimator.
\newpage
%
\small
\bibliographystyle{ieeetr}
\bibliography{/Users/nico/Documents/Library/nicolib.bib}
%
%
%
\end{document}